\documentclass[floatfix, preprintnumbers, twocolumn, 10pt]{revtex4}
\usepackage{amsmath,amsfonts,amssymb,graphicx}
\usepackage{dcolumn}
\usepackage{bm}
\usepackage{epsfig}
\usepackage{epstopdf}

\def\bea{\begin{eqnarray}}
\def\eea{\end{eqnarray}}
\def\be{\begin{equation}}
\def\ee{\end{equation}}
\def\ba{\begin{array}}
\def\ea{\end{array}}

\begin{document}

\newcommand{\eq}[2]{\begin{equation}\label{#1}{#2}\end{equation}}
\date{\today}
\preprint{HU-EP- 09/18}
\title{UV/IR mode mixing and the CMB}

\author{Gonzalo A. Palma$^{1),2)}$}  \email[email: ]{palma@lorentz.leidenuniv.nl}
\author{Subodh P. Patil$^{3)}$}  \email[email: ]{subodh@physik.hu-berlin.de}

\affiliation{1) Instituut-Lorentz for Theoretical Physics, Universiteit Leiden, NL-2300 RA Leiden, The Netherlands}
\affiliation{2) Physics Department, FCFM, Universidad de Chile, Blanco Encalada 2008, Santiago, Chile}
\affiliation{3) Humboldt Universit\"at zu Berlin, Institut f\"ur Physik, Newtonstr. 15, D-12489 Berlin, Germany}

\begin{abstract}
It is well understood that spatial non-commutativity, if indeed realized in nature, is a phenomenon whose effects are not just felt at energy scales comparable to the non-commutativity scale. Loop effects can transmit signatures of any underlying non-commutativity to macroscopic scales (a manifestation of a phenomenon that has come to be known as UV/IR mode mixing) and offer a potential lever to constrain the amount of non-commutativity present in nature, if present at all. Field theories defined on non-commutative spaces (realized in string theory when D-branes are coupled to backgrounds of non-trivial RR background flux), can exhibit strong UV/IR mode mixing, manifesting in a non-local one loop effective action. In the context of inflation in the presence of any background non-commutativity, we demonstrate how this UV/IR mixing at the loop level can allow us to place severe constraints on the scale of non-commutativity  if we presume inflation is responsible for large scale
  structure. We demonstrate that any amount of non-commutativity greatly suppresses the CMB power at all observable scales, {\it independent} of the scale of inflation, and independent of whether or not the non-commutativity tensor redshifts during inflation, therefore nullifying a very salient and successful prediction of inflation.
\end{abstract}
\maketitle

\section{Introductory Remarks}

The notion that the fundamental structure of spacetime might be something other than a continuum has been around for many decades. A particular type of modified spacetime structure, which in modern parlance has come to be known as non-commutative geometry, had in fact been studied as far back as the 1940's \cite{snyder}. Recent developments in string theory suggest that this modification to geometry at high energy scales can arise in certain D-brane backgrounds with a non-trivial RR flux, resulting in a resurgence of interest in the phenomenological consequences of physics on non-commutative spaces.

The basic idea of non-commutative geometry is rather simple: Instead of considering spacetime coordinates  $x^{\mu}$ (with $\mu = 0,\ldots,3$) to be described by c-number functions on some manifold,  we consider our spacetime coordinates to satisfy some sort of structure relations
\eq{nc}{[\hat x^\mu,\hat x^\nu] = i O^{\mu\nu},}
where the $\hat x^\mu$ are Hermitian operators, and $O^{\mu\nu}$ is a Hermitean operator antisymmetric in its indices. Several widely studied examples include the type of non-commutativity first introduced in \cite{snyder}
\eq{ncg}{[\hat x^\mu,\hat x^\nu] = i I\theta^{\mu\nu},}
where $I$ is the identity operator and $\theta^{\mu\nu}$ is a non-degenerate antisymmetric tensor, and
\eq{fuzz}{[\hat x^i,\hat x^j] = i\kappa \epsilon_{ijk}\hat x^k,}
(with $i = 1,2,3$ denoting spatial components) where $\kappa$ has dimensions of length, and corresponds to the scale of non-commutativity. Both examples arise in string theory when describing D-brane dynamics in non-trivial flux backgrounds \cite{vs, myers}. In this report, we focus on the former relation (\ref{ncg}), and explore its consequences for inflationary cosmology, where we will find that we can tightly constrain the amount of spacetime non-commutativity present  during inflation.

The initial intuition that motivated this study, was that the UV/IR mode mixing present in theories such as those defined on non-commutative (Moyal) spaces defined by~(\ref{ncg}), could provide us a constraining lever on physics at high energy scales complimentary to the lever offered by the effect of modes being stretched by inflation \footnote{The latter was the motivation of many investigators concerned with the so-called `trans-Planckian' problem (see~\cite{rhb} for a review).}. What we find is that the presence of this effect present in this example is so severe as to already be ruled out by present observations (if it hadn't already been severely constrained for other reasons~\cite{rh, abel}). In this sense, the spirit in which we undertake this study is that of a first pass (within a concrete and widely studied context), on the notion that UV/IR mode mixing may offer us a complimentary lever on very high energy physics in the CMB, which in this example, is used to rule
  out a class of modifications to space time structure at such energy scales. 

We begin by reviewing those aspects of non-commutative field theory relevant for our study, after which we will consider inflation in the presence of background non-commutativity. After showing that the background inflaton dynamics are identical for the commutative vs the noncommutative cases (in the event that non-commutativity is purely spatial), we derive the action for the scalar perturbations, where we find that loop effects render a non-local effective action. We show that the net effect of spatial non-commutativity is to modify the dispersion relations for the propagation of the perturbations, such that mode propagation differs from the usual case for all wavelengths {\it less} than the critical wavelength $k_c = (\epsilon H/M_{pl})^{1/4}/l_\theta$, where $l_\theta$ is the non-commutativity length scale, and $\epsilon$ is the slow roll parameter. We see that UV/IR mode mixing induced by loop effects communicates signatures of non-commutativty to scales far below the no
 n-commutativity scale independent of whether or not the theta tensor redshifts during inflation. We will then compute the two point correlation functions for the metric fluctuations, and find that the presence of any amount of spatial non-commutativity to be inconsistent with the observed CMB spectrum.  

\section{Preliminaries}

In what follows, we offer a brief overview of those aspects of non-commutative field theory relevant for our investigation. We refer to the excellent review of ref.~\cite{nd} for details not elaborated upon here. To start with, once we assume the structure relations $[\hat x^\mu,\hat x^\nu] = i I\theta^{\mu\nu}$ for our spacetime coordinates, we find that the net effect of spacetime non-commutativity is to deform the usual product on our manifold to the so-called star product, defined by
\eq{sp}{f\cdot g(x) \to f*g(x) \equiv e^{\frac{i\theta^{\mu\nu}}{2}\frac{\partial}{\partial x^\mu}\frac{\partial}{\partial y^\nu}}f(x) g(y)|_{y=x}.}
For readers unfamiliar with this aspect of non-commutative physics, we offer a quick review of why this is so in appendix \ref{app-A}. Field theories on non-commutative spaces can be defined in the usual way, except that now all products in the action are to be replaced with star products
\eq{efact}{S = -\int d^4 x~[ \frac{1}{2} \partial_\mu \phi * \partial_\mu \phi + V_*(\phi)],}
where $V_*(\phi)$ indicates that all fields are multiplied with star products. It is a property of the star product that the quadratic part of any action remains unaffected   
\eq{inteq}{\int d^4x~ f\cdot g = \int d^4x~ f*g,}
as is easily checked by Fourier expanding $f$ and $g$ and performing the spacetime integral. Therefore the quadratic part of the action is unaffected by the underlying non-commutative structure. In particular, this means that the Green's functions for the commutative and the non-commutative theories are identical, and suffer from the same short distance divergences \cite{filk}. Consider now the $\phi^4$ theory in four dimensions:
\eq{p4a}{S = - \int d^4 x \Bigl[\frac{1}{2}(\partial\phi)^2 + \frac{m^2}{2}\phi^2 + \frac{g^2}{4!}\phi*\phi*\phi*\phi \Bigr].}
As we see immediately, the underlying non-commutative structure affects only the interaction terms. Working in Fourier space, it is straightforward to show that the net effect of non-commutativity on the interaction vertices is to introduce an additional phase factor relative to the commutative case \cite{filk}
\eq{intv}{V(k_1,..,k_n) = e^{-\frac{i}{2}\sum_{i<j} k_i\times k_j} ,}
where $k_i$ is the $i^{\rm th}$ momentum flowing into the vertex, and we employ the shorthand notation:
\eq{shn}{k_i\times k_j = k_{i\mu}\theta^{\mu\nu}k_{j\nu}.}
One should note that $V(k_1,..,k_n)$ is not invariant under arbitrary permutations of the $k_i$, but only cyclic permutations. This results in a relative phase factor between the two one loop self energy diagrams shown in FIG.~1. 
\begin{figure}[tbh]
\epsfig{file=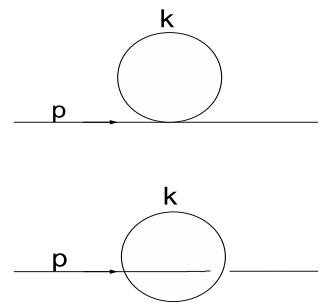, height=2in, width=1.8in} 
\caption{Planar and non-planar self energy graphs for $\phi^4$ theory.}
\label{fd}
\end{figure}
These two diagrams make the 1-loop 1PI two point function to differ by a phase of the form \cite{min}
\begin{eqnarray}
\Gamma_{\rm 1\,planar} &=& \frac{g^2}{3(2\pi)^4}\int \frac{d^4k}{k^2 + m^2},\\
\Gamma_{\rm 1\,nplanar} &=& \frac{g^2}{6(2\pi)^4}\int \frac{d^4k}{k^2 + m^2}e^{ik\times p},
\end{eqnarray}
where they would have yielded identical contributions in the commutative case (up to symmetry factors). Following the treatment of ref.~\cite{min}, we observe that non-commutativity appears to regulate the non-planar sector of the theory (through the oscillating phase factor), whilst leaving the planar sector unaffected. In order to see the effects of this phase factor, we use the Schwinger reparameterization of the propagator
\eq{sr}{\frac{1}{k^2 + m^2} = \int^\infty_0 d\alpha e^{-\alpha(k^2 + m^2)},}
which renders the loop momentum integrations Gaussian. We can evaluate these straight away to yield
\begin{eqnarray}
\Gamma_{\rm 1\,planar} &=& \frac{g^2}{48\pi^2}\int \frac{d\alpha}{\alpha^2}e^{-\alpha m^2},\\
\Gamma_{\rm 1\,nplanar} &=& \frac{g^2}{96\pi^2}\int \frac{d\alpha}{\alpha^2}e^{-\alpha m^2 -\frac{p \circ p}{\alpha}},
\end{eqnarray}
where we adopt the notation:
\eq{pop}{p \circ k = -p_\mu\theta^{\mu}_{\,\,\, \nu} \theta^{\nu\lambda}k_\lambda.}
Regulating the small $\alpha$ divergence with the cut-off $\exp[-1/(\Lambda^2\alpha)]$ yields
\begin{eqnarray}
\Gamma_{\rm 1\,planar} &=& \frac{g^2}{48\pi^2}\int \frac{d\alpha}{\alpha^2}e^{-\alpha m^2 -\frac{1}{\Lambda^2\alpha}},\\
\Gamma_{\rm 1\,nplanar} &=& \frac{g^2}{96\pi^2}\int \frac{d\alpha}{\alpha^2}e^{-\alpha m^2 -\frac{p \circ p + 1/\Lambda^2}{\alpha}},
\end{eqnarray}
from which we can immediately read  that if $\Lambda$ is the momentum cut-off for the planar sector, then the non-planar sector has a cut-off:
\eq{effc}{\Lambda^2_{\rm eff} = \frac{1}{1/\Lambda^2 + p\circ p}.}
In the event that not all coordinates are non-commutative, $p \circ p$ is to be understood as being only constructed out of the non-commutative components of the 4 momenta. We evaluate the contributions to the one loop effective actions as:
\begin{eqnarray}
\Gamma_{\rm 1\,planar} &=& \frac{g^2}{48\pi^2}\Bigl( \Lambda^2 - m^2 \ln \frac{\Lambda^2}{m^2} +... \Bigr) ,\\
\Gamma_{\rm 1\,nplanar} &=& \frac{g^2}{96\pi^2}\Bigl( \Lambda_{\rm eff}^2 - m^2 \ln \frac{\Lambda_{\rm eff}^2}{m^2} +... \Bigr).
\end{eqnarray} 
We note from the form of eq.~(\ref{effc}) that in the limit $\Lambda \to \infty$, the non-planar contribution to the effective action diverges in the limit $p \to 0$. That is to say, the non-planar sector appears to inherit an IR divergence at low momenta from the UV divergence of the planar sector. In momentum space, the resulting effective action from the contributions of all self energy graphs is given by
\begin{eqnarray}
S &=& \int d^4p \Bigl( p^2 + M^2 + \frac{g^2}{96\pi p\circ p } \nonumber \\ 
&& \quad\quad - \frac{g^2 M^2}{96\pi} \ln  \frac{1}{M^2p\circ p}  \Bigr)\phi(-p)\phi(p), \label{1lea}
\end{eqnarray}
where $M^2$ is the renormalized mass given by:
\eq{mren}{M^2 = m^2 + \frac{g^2\Lambda^2}{48\pi^2} - \frac{g^2 m^2}{48\pi^2} \ln \frac{\Lambda^2}{m^2} .}
Thus we see from eq.~(\ref{1lea}), that loop effects can communicate even high scale non-commutativity to large wavelengths, as is evidenced by the non-local term proportional to the self coupling, which becomes significant for low momenta.

We take note of a similar effect for $\phi^3$ theory in 4-d. If we denote the (now dimensionful) coupling constant as $g$, we see from FIG.~2 that again there is a planar and a non-planar sector to the one loop self energy. 
\begin{figure}[tb]
\epsfig{file=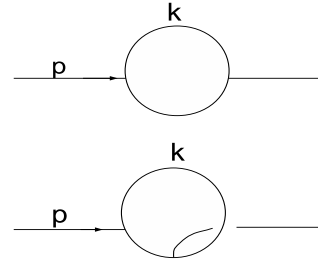, height=2in, width=1.8in} 
\caption{Planar and non-planar self energy graphs for $\phi^3$ theory}
\label{fd2}
\end{figure}
It follows for similar reasons that the planar and the non-planar sectors contribute differently to the one loop effective actions and yield \cite{nd}
\eq{1lea3}{S = \int d^4p \Bigl( p^2 + M^2 - \frac{g^2}{256\pi^2} \ln \frac{1}{M^2p\circ p} \Bigr)\phi(-p)\phi(p),}
where now, unlike the case of $\phi^4$ theory, the leading singular term at low momenta for the one loop effective action for $\phi^3$ theory is only logarithmically divergent. We note in passing that although the IR divergent terms in the low energy effective action were derived in the context of imposing a cut-off regularization and sending this cut-off to infinity, the IR divergence derived above are to be taken to be as generic features of theories which exhibit strong UV/IR mixing. In fact, in \cite{min} it was shown in the context of $\phi^4$ theory that quadratic IR divergences result independent of the scale of the cut-off $\Lambda$ \footnote{We also note that although derived here in a perturbative context, this UV/IR mixing has been seen to manifest non-perturatively as well \cite{panero}.}.

In the next section, we investigate the consequences of such loop effects (UV/IR mixing) for inflation in the presence of any underlying non-commutativity, and compute two point functions for the scalar fluctuations.

\section{Inflation and non-commutativity}

We now commence our analysis of the effects of non-commutativity on CMB observables. First, we posit that there be no temporal non-commutativity ($\theta^{0i} =0$) and consider
\be
[\hat x^i, \hat x^j] = i\theta^{ij} ,  \label{spat-non-commutativity}
\ee
(with $i = 1,2,3$) which is necessary if we are to preserve unitarity \cite{gomis} as well as requiring that our construction be consistently realized in string theory.
With this in mind, it is possible to see that the dynamics for homogeneous field configurations $\phi = \phi(t)$ remain unaffected by non-commutativity. In particular, in a Friedmann-Robertson-Walker (FRW) background
\be
ds^2 = - dt^2 + a^2(t) \delta_{ij} dx^i dx^j, \label{FRW-metric}
\ee
 the equations of motion $\square\phi - V_*'(\phi) = 0$ reduce to the usual commutative equations (with $H=\dot a/a$)
\eq{eomsc}{\ddot \phi + 3H \dot \phi + V'(\phi) = 0 ,}
for homogeneous field configurations $\phi = \phi(t)$, given eq.~(\ref{sp}) and the fact that $\theta^{0i} = 0$:
\eq{homeq}{\phi(t)*\phi(t) = \phi(t)\cdot\phi(t).}
An alternative way to see this is to consider the corresponding Weyl operator equations of motion (see appendix \ref{app-A})
\be
\ddot{\hat\Phi} + 3H\dot{\hat\Phi} + V'(\hat\Phi) = 0,
\ee
and realizing that any commutative homogeneous solution $\phi(t)$ is a solution at finite theta by considering the matrix solutions $\hat\Phi = \phi(t) \hat I$. 
Hence given any purely spatial non-commutativity during inflation, we will have the usual solutions for the background.

Inflaton perturbations $\varphi(t,x) = \phi(x,t) - \phi_0(t)$, on the other hand, will feel the underlying non-commutativity and as shown in appendix B, will result in the following action for the comoving curvature perturbations $\zeta = - \varphi / \sqrt{2 \epsilon}$:
\begin{eqnarray}
\label{pertac}S &=& M^2_{pl}\int \sqrt{-g}~d^4x \epsilon~\Bigl[ \zeta \square\zeta + \zeta\frac{\lambda}{\partial\circ\partial}\zeta\Bigr]
\end{eqnarray}
where $\partial \circ \partial = - \theta^{i k} \theta_{k}^{\,\, j} \partial_{i} \partial_{j}$ involves 
only spatial derivatives and $\lambda$ is a dimensionless parameter given by:
\eq{lamb}{\lambda =  c \frac{H^2}{M^2_{pl}} \delta^2 .}
where $\delta$ is taken to be commensurate to the slow roll parameters $\epsilon$ and $\eta$, and its precise value will depend on the details of the background solution. 
In the previous expression $c$ is a number of order unity which depends on the specifics of the underlying inflationary model, whereas $\epsilon$ and $\eta$ are the usual slow roll parameters:
\eq{srp}{\epsilon = \frac{\dot\phi_0^2}{2 H^2 M_{pl}^2}, \qquad \mathrm{and} \qquad \eta = -\frac{\ddot \phi_0}{H \dot \phi_0}.} 
The action of eq.~(\ref{pertac}) is derived in Appendix~B [eq.~(\ref{pertactss})] and includes one-loop effects arising from non-commutativity as a consequence of higher order interactions in the inflaton perturbations proportional to $\epsilon$.

At short wavelengths, the order of non-commutative effects comes determined by the size of the  lengthscale $l_{\theta}$, entering the non-commutative structure tensor $\theta^{ij} \sim l_{\theta}^2$. It is rather clear that the validity of slow-roll inflationary framework in a non-commutative background requires us to have 
\be
l_{\theta} H \ll1,   \label{cond-desitter}
\ee
which simply means that the de Sitter radius $H^{-1}$ determining the size of causally connected regions during inflation has to be much larger than the non-commutative lengthscale $l_{\theta}$.
We also note here that eq.~(\ref{pertac}) is a one loop effective action, which diverges at low momenta (and where higher loop effects become significant), has a domain of validity which does not stretch arbitrarily far into the IR. As demonstrated in \cite{min}, this domain of validity is given by
\eq{domval}{m^2 \, p\circ p \geq \mathcal O (e^{-\frac{\tilde c M^2_{pl}}{\delta^2 H^2}}),} 
where $\tilde c$ is of order unity, and as shown in the appendix, $m^2 \sim \delta H^2$. For any reasonable model of inflation, both $H/M_{pl}$ and  $\delta$ are smaller than $1/100$, therefore making the exponent on the right hand side almost vanishing, and certainly all scales of cosmological interest fall within the domain of validity of our approximation. 

\section{Non-Commutativity and the CMB}

We now study the evolution of perturbations during inflation in the presence of non-commutativity, and its consequences for the spectrum of scalar fluctuations. In what follows, we rewrite the metric of eq.~(\ref{FRW-metric}) and work in conformal coordinates
\eq{ccs}{ds^2 = a^2(\eta) (- d\eta^2 + \delta_{ij} dx^i dx^j),} 
where $\eta$ is the usual conformal time satisfying $d\eta = a^{-1} dt$. Furthermore, to simplify our analysis, we will consider the evolution of perturbations on a de Sitter background $a(\eta) = -1/ H \eta$, where $H^{-1}$ corresponds to the de Sitter radius, and consider the approximation whereby $H$ evolves adiabatically. From eq.~(\ref{pertac}) the equation of motion for the curvature perturbations is found to be \footnote{We note that the non-local term is easily accounted for in the action principle, as with the choice $\theta^{0i}=0$ and the FRW form of the metric tensor (\ref{ccs}), the operator $\partial\circ\partial$ is self-adjoint with respect to the scalar product $(\phi,\psi) = \int \sqrt{-g}\phi~\psi$, and can hence be brought over to act on the partner field in any bi-linear product. (\ref{peom}) Then follows directly.}
\eq{peom}{\square\zeta + \frac{\lambda}{\partial\circ\partial}\zeta = 0.}
By using the metric of eq.~(\ref{ccs}) and realizing that $\theta^{i}_{\,\,\, j}$ is purely spatial, we can write the above in momentum space:
\be
\zeta'' -\frac{2}{\eta}\zeta' + \Bigl(k^2 + \frac{\lambda}{H^2 \eta^2 \theta^{i}_{\,\,\, j} \theta^{j}_{\,\,\, m} k^{m} k_{i} }\Bigr)\zeta = 0. \label{mode-eq}
\ee
Since $\theta^{ij}$ has to have even rank \cite{nd, min}, we pick two directions in which $\theta^{xy} = -\theta^{xy}$. That this choice breaks isotropy is inevitable. In \cite{bal, Akofor:2008gv} the consequences of this preferred direction for the CMB was explored, and in \cite{shiu, china, Lizzi:2002ib}, its consequences for non-gaussian imprints in the CMB were considered. We wish to show in this report is that if indeed non-commutativity is present, these effects are subsidiary, and are dwarfed by the fact that CMB power is greatly suppressed at long wavelengths by the effects of non-planar UV/IR mixing. That is, we are not concerned with these aspects of what we view as model specific features. Instead, we are more interested in exploring what we view as a generic consequences of UV/IR mixing in the context of a concrete model, and in doing so, deriving constraints on the presence of non-comutativity before, or during inflation.

To make our analysis as exhaustive as possible, in this report we consider various possible time dependences of the tensor $\theta^{ij}$ at the right hand side of eq.~(\ref{spat-non-commutativity}). To be more precise --and without loss of generality-- we will take the coordinates $x^{i}$ in eq.~(\ref{spat-non-commutativity}) to coincide with the co-moving coordinates of the metric in eq.~(\ref{ccs}) and consider, separately, the three different possibilities whereby either $\theta^{ij}$, $\theta^{i}_{\,\, j}$ or $\theta_{ij}$ are constant tensors (i.e. independent of the metric), and indices are raised with the spatial part of the metric $g^{ij} = a^{-2}(\eta) \delta^{i j}$. 

In the context of string theory, different time dependences of $\theta^{ij}$ can be traced back to different realizations of inflation.  Supposing, for instance, that the inflaton is a transverse modulus describing the end points of open strings (D-branes) in non-trivial $B$-field backgrounds, the open string endpoints (brane coordinates) will satisfy the Moyal type non-commutative structure relations \cite{nd}:
\eq{stncg}{\theta^{ij} =  \alpha'\Bigl(\frac{1}{g + \alpha' B}\Bigr)^{ij}_A       .}
where $g$ is the string worldsheet metric and $B$ is the worldsheet anti-symmetric field (with dimensions of inverse length squared), and $A$ denotes anti-symmetrization. We note that expanding in powers of $\alpha'$, to leading order, two factors of the metric will contract $B_{ij}$, leading to $\theta^{i j} = - \alpha'^{2}g^{i m}g^{j n} B_{m n} + \mathcal{O}(\alpha'^3)$, and thus we see that in the particular case where the background is characterized by a constant $B_{ij}$ field, we recover a constant $\theta_{ij}$. One can imagine, however, a situation in which inflation is being realized as the result of a moving D-brane on a warped background with $B_{ij}$  scaling spatially accordingly. In such a case, the effective time variation of $B_{ij}$ on the brane would be such that $\theta^i_{\,\, j}$ (instead of $\theta_{ij}$) remains constant. Independently of the details of how inflation is actually realized, in what follows consider the three aforementioned cases.


\subsection{$\theta^{ij}$ independent of the metric tensor}

The first case of interest consists of the choice in which $\theta^{ij}$ is independent of the metric. By writing $\theta^{xy} = -\theta \epsilon^{xy}$ we then obtain
\be
\theta^{i}_{\,\,\, j} \theta^{j}_{\,\,\, m} k^{m} k_{i} = a^{2} \theta^2 \bar k^2 = \frac{\theta^2 \bar k^2}{H^2 \eta^2}, \label{theta-case-1}
\ee
where $\bar k^2 = k_x^2 + k_y^2$ is the momentum squared along the non-commuting directions. Notice that we can rewrite the non-commutative relation in terms of {\it proper} coordinates $\hat y^i \equiv a(t) \hat x^i$ in the following way:
\be
[\hat y^i, \hat y^j] =   i a^{2}(\eta) \theta^{ij} .
\ee
This means that during inflation the proper noncommutative length scale $l_{\theta}$ grows with time as $l_{\theta} \equiv a(\eta) l_{0} $ where $l_0 \equiv \sqrt{\theta}$ is a co-moving length scale. By inserting eq.~(\ref{theta-case-1}) in eq.~(\ref{mode-eq})
we arrive to the following equation of motion:
\eq{mfeq}{\zeta'' -\frac{2}{\eta}\zeta' + \Bigl(k^2 + \frac{\lambda}{\theta^2\bar k^2}\Bigr)\zeta = 0.}
We immediately read off the solutions to this equation, as they only differ from the standard canonically normalized mode functions in their $k$ dependence
\eq{mfsolns}{\zeta_{\pm} = \frac{H}{\sqrt{2\epsilon K^3}}(1 \mp iK\eta)e^{\pm i (K\eta - \vec k\cdot x)},}
with $K$ defined as
\eq{kdef}{K = \sqrt{\vec k^2 + \frac{\lambda}{\theta^2 \bar k^2}}.}
We take note of the following duality in the above
\eq{dual}{k_x^2 + k_y^2 = \bar k^2 \leftrightarrow \frac{\lambda}{\theta^2 \bar k^2},}
which has a simple interpretation in terms of the scattering of localized $\varphi$ wave packets upon self-interaction \cite{min}. We can immediately see how this will lead to an unacceptable modification to the CMB spectrum, namely, that any such duality will cause the CMB to be symmetric around the pivot wavelength:
\eq{kpiv}{\bar k_p = \lambda^{1/4}/l_\theta.}
Note that even in the case of very high scale non-commutativity, even as high as Planckian, this pivot wavelength will be below this scale (for $\epsilon$ and $H/M_{pl} \sim 1/100$, this scale becomes the GUT scale). Any feature of the spectrum at short wavelengths will be exactly mirrored at long wavelengths. Hence, if the normalization of the mode functions vanishes towards the UV, it will also do so in the IR. This will result in suppression of power at low multipoles, as is easily seen by using standard techniques to compute the resulting power spectrum for curvature perturbations (defined in Appendix \ref{app-B})
\eq{cp}{\langle \zeta(\vec k)\zeta(\vec k') \rangle = (2\pi)^3\delta^3(\vec k - \vec k')\frac{H^2}{2K^3\epsilon^2 M_{pl}},}
which results in the power spectrum:
\eq{ps}{\mathcal P(k) = \frac{1}{\epsilon}\frac{H^2}{2\pi M^2_{pl}}\frac{k^3}{K^3} = \frac{1}{\epsilon}\frac{H^2}{2\pi}\frac{k^3}{( k^2 + \lambda/\theta^2 \bar k^2)^{3/2}}.}
We easily read off that the two point function will start to receive significant corrections {\it below} the wavlength (\ref{kpiv}). If we impose COBE normalization at the comoving wavelength $k_C$ via observational input, the above implies in order for non-commutativty to not affect this normalization, that
\eq{ncs}{k_\theta  \ll k_C/\lambda^{1/4},}
in other words, non-commutativity at cosmological scales, a clearly untenable state of affairs. Viewed from another perspective, if non-commutativity were present at some high energy scale before, or during inflation, the resulting spectrum of metric fluctuations would have vanishing power at all observable scales-- that is, the main feature of inflation that inspires our belief in it would be nullified by non-commutativity and we'd have to search elsewhere for the origins of large scale structure. This `stiffening' of long wavelength fluctuations is a well known and widely studied feature of non-commutative field theories (see \cite{gubser} and references therein), and we see here that in the context of inflation, would nullify the mechanism with which it generates the primordial fluctutations responsible for large scale structure.

If the reader as doubts at this point that this may be an artifact of the fact that we considered the case where the theta tensor is independent of the metric tensor, in the next section we demonstrate that this stiffening of long wavelength modes is even more severe in the opposite case.


\subsection{$\theta^{i}_{\,\,\, j}$ independent of the metric tensor}

The second case of interest corresponds to the situation in which $\theta^{i}_{\,\,\, j}$ (with both upper and lower indices) is  independent of the metric. Notice that in this case one has
\be
[\hat x^i, \hat x^j] = i g^{i k} \theta_{k}^{\,\,\, j} = i a^{-2} (t) \theta_{i}^{\,\,\, j},
\ee
and therefore the non-commutative relation expressed in {\it proper} coordinates $\hat y^i \equiv a(t) \hat x^i$ is given by:
\be
[\hat y^i, \hat y^j] =   i \theta_{i}^{\,\,\, j} ,
\ee
which means that the proper noncommutative length scale $l_{\theta} \equiv \sqrt{\theta}$ remains constant during the inflationary stage. If, similarly to the previous case, we write $\theta_{x}^{\,\,\, y} = - \theta \epsilon^{x y}$, we are then left with $\theta^{i}_{\,\,\, j} \theta^{j}_{\,\,\, m} k^{m} k_{i} = H^2 \eta^2 \theta^2 \bar k^2$ and the equations of motion~(\ref{mode-eq}) for the mode functions take the form
\be
\zeta'' -\frac{2}{\eta}\zeta' + \Bigl(k^2 + \frac{\lambda}{\eta^4 H^4 \theta^2 \bar k^2 }\Bigr)\zeta = 0. \label{second-case-eq}
\ee
Notice that the term arising from the 1-loop contribution becomes dominant as $\eta \rightarrow 0$. It is rather clear that this term cannot dominate the dynamics for modes at any stage during inflation, if these are to be held responsible for the CMB anisotropies observed today. To appreciate this, let us suppose that this term indeed dominates. Then, in the long wavelength limit, the canonically normalised solution to the equations are found to be
\be
\zeta_{\pm} = \frac{H \eta^2}{\sqrt{\epsilon}C^{1/2}(k)} e^{\mp \frac{ i C(k)}{\eta}} e^{\pm \vec k \cdot \vec x} , \label{second-case-sol}
\ee
where $C(k)$ is given by
\be
C^2(k) = \frac{\lambda}{H^4 \theta^2 \bar k^2}.
\ee
From this, one can immediately evaluate the two point correlation function of curvature perturbations $\zeta(\vec k)$, which reads:
\be
\label{2pcc2}
\langle \zeta(\vec k)\zeta(\vec k') \rangle = (2\pi)^3 \delta^3(\vec k - \vec k')  \frac{H^2 \eta^4}{C(k)M^2_{pl}\epsilon}.
\ee
At this point, it should be clear that the power spectrum deduced out of this quantity is strongly dependent on the scales $\bar k$ via $C(k)$, loosing one of the most important features found in slow-roll inflation, namely the prediction of a nearly scale invariant power spectrum. Furthermore, we note from (\ref{second-case-eq}) that the curvature perturbation no longer ceases to evolve after horizon crossing. Instead, it is damped until the end of inflation, after which non-commutative effects are no longer felt and the curvature perturbation proceeds conserved until it re-enters our horizon as imprints on the CMB. Hence the observed power spectrum, on top of its shape being highly distorted, is extremely suppressed:
\eq{psc2}{ P(k) = \frac{1}{\epsilon}\frac{H^2}{2\pi M^2_{pl}}\Bigl[H^2\theta e^{-4N(k)}\Bigr],}
where $N(k)$ is the number of e-folds the mode $k$ undergoes after it exits the horizon until inflation ends, whence it is the appearance of the factors of $\eta$ in (\ref{2pcc2}) that result in this drastic suppression (and blueshifting of the spectrum).

It is important to notice that, as $\eta \rightarrow 0$, the momenta of the mode solutions is redshifted until eq.~(\ref{domval}), determining the validity of this regime, is no longer satisfied. At this point, higher loop contributions start kicking in, modifying the asymptotic solution expressed in eq.~(\ref{second-case-sol}). It is nevertheless clear that these higher order loop effects will not restore the scale invariance of the power spectrum.

We can in fact deduce a strong constraint on the parameter $\theta$ by requiring that 1-loop contributions to the equation of motion (\ref{second-case-eq}) do not affect the evolution of modes presently observed at the CMB. For this, let us consider those modes observed in the CMB with the longest wavelengths. Roughly speaking, these modes crossed the horizon about $N \sim 55$ $e$-folds before the end of inflation, the precise number depending on the specific model under consideration. As we have seen, 1-loop corrections can drastically affect the evolution of these modes outside the horizon, and therefore the term $ \frac{\lambda}{\eta^4 H^4 \theta^2 \bar k^2 }$ in eq.~(\ref{second-case-eq}) has to remain suppressed during the whole $N \sim 55$ $e$-folds following the horizon crossing $k \sim  \eta^{-1}$. In order to get an estimate of how suppressed this term has to be, let us consider the equations of motions~(\ref{second-case-eq}) in terms of the re-escaled field $\chi \equiv a(\eta) \zeta$
\be
\chi'' + \Bigl(k^2 - \frac{2}{\eta^2} + \frac{\lambda}{\eta^4 H^4 \theta^2 k^2 }\Bigr)\chi = 0, \label{second-case-eq-chi}
\ee
where, in order to obtain a conservative bound, we have for simplicity replaced $\bar k^2 \rightarrow k^2$. 
Then, the requirement of having a suppressed 1-loop contribution affecting the evolution of modes during the last 55 $e$-folds after horizon crossing $k^2 \sim \eta^{-2}$ is given by
\be
\frac{|\lambda|}{H^4 \theta^2} < e^{- 2 N}, \label{bound-1}
\ee
where $N \sim 55$. To proceed, let us recall that $\lambda \simeq \delta^2 H^2 / M_{pl}^2$. Moreover, as shown in the Appendix B, there is at least one contributing term in $\lambda$ of order $\epsilon \eta H^2 / M_{pl}^2$, which for the present discussion we treat as the leading piece \footnote{Current observations constrain the linear combination $n_s - 1 = 2 (\eta - 2 \epsilon)$ to be of order $\sim 0.04$ \cite{Komatsu:2008hk}, therefore implying that $\eta$ is at least of that same order (recall that $\epsilon>0$). $\epsilon$ on the other hand is not well constrained yet and might be much smaller than $\sim 0.01$.}. Then,  since the spectral index is given in terms of the slow roll parameters as $n_s - 1 = 2 (\eta - 2 \epsilon)$, we can therefore write $|\lambda| \simeq\epsilon |n_s - 1| H^2 / 2 M_{pl}^2$. Moreover, provided that eq.~(\ref{bound-1}) holds, one can use the COBE normalization of the spectrum and write $H \simeq 10^{-4} \epsilon^{1/2} M_{\rm Pl}$ to eliminate $\epsilon$ in terms of $H$. Putting all of this together, we then obtain:
\be
l_{\theta}^{-1} < 10^{-2} e^{-N/2} |n_s-1|^{-1/4} M_{pl}.
\ee
Given that the spectral index is measured to be $n_s \sim 0.96$ \cite{Komatsu:2008hk}, we obtain the following bound on the non-commutative lengthscale:
\eq{ltconst}{l^{-1}_\theta < 10^{-14}M_{pl} \sim 10^4 \mathrm{GeV}.}
That is, the non-commutative energy scale is required to be close to current high energy experiments. Moreover, from eq.~(\ref{cond-desitter}) we  see that  the inflationary scale had to be well within reach of current observations.


\subsection{$\theta_{ij}$ independent of the metric tensor}

Finally, consider now the non-commutative relation (\ref{mode-eq}) in the case where $\theta^{ij} = g^{i m} g^{j n} \theta_{mn} = a^{-4}(\eta) \theta_{ij}$ with $\theta_{ij}$ a constant tensor. As before, we introduce $\theta$ by writing $\theta_{xy} = -\theta \epsilon_{xy}$, which allows us to deduce:
\be
\theta^{i}_{\,\,\, j} \theta^{j}_{\,\,\, m} k^{m} k_{i} = H^6 \eta^6 \theta^2 \bar k^2, \label{theta-case-2}
\ee
Notice that in this case the proper coordinates $\hat y^i \equiv \hat a(\eta) x^i$ satisfy the following non-commutative relation
\be
[\hat y^i, \hat y^j] =   i a^{-2}(\eta) \theta_{ij} ,
\ee
and therefore the proper non-commutative lengthscale decreases during inflation as $l_{\theta} = a^{-1}(\eta)  l_0$, where $l_0 = \sqrt{\theta}$. Interestingly, even though this lengthscale is becoming smaller as inflation progresses, due to the way in which IR effects are accentuated by inflation, a non-zero $l_0$ produces drastic modifications of the spectrum, as we see in the following. First, by replacing eq.~(\ref{theta-case-2}) in eq.~(\ref{mode-eq}) we obtain
\eq{mfeq2}{\zeta'' -\frac{2}{\eta}\zeta' + \Bigl(k^2 + \frac{\lambda}{\theta^2\bar k^2 H^8\eta^8}\Bigr)\zeta = 0.}
As before, if the 1-loop contribution dominates ($H\eta \to 0$), we can ignore the first term to obtain the canonically normalized solutions
\eq{msu}{\zeta_\pm = \frac{H \eta^3}{\sqrt{2\epsilon} C^{1/2}(k)}e^{\mp\frac{i C(k)}{3\eta^3}}e^{\pm i \vec k \cdot \vec x},} 
with $C(k)$ (which has dimensions inverse mass cubed) is defined by:
\eq{ck}{C^2(k) = \frac{\lambda}{\theta^2 H^8 \bar k^2}.}
From this, one can immediately evaluate the two point correlation function in the long wavelength limit to find:
\be
\label{fc2p}
\langle \zeta(\vec k)\zeta(\vec k') \rangle = (2\pi)^3 \delta^3(\vec k - \vec k')  \frac{H^2 \eta^6}{2M^2_{pl}\epsilon C(k)}.
\ee
As before, we see that power spectrum deduced from this expression is not only heavily suppressed, but it is heavily dependent on the scales due to the scale dependence posed by $C(k)$. We can now repeat the previous analysis and ask ourselves what the value of $\theta$ should be in order to avoid modifications to the evolutions of modes observed at the CMB after they left the horizon. To this extent, notice that since the proper lengthscale $l_{\theta} = a^{-1}(\eta) l_0$ is varying with time, then $l_0= \sqrt{\theta}$ determines the value of $l_{\theta}$ at the time $\eta_0$ where $a(\eta_0) = 1$, which we are free to specify. Let us choose $\eta_0$ to correspond to the epoch in which the longest wavelengths present in the CMB crossed the horizon [recall the discussion immediately before eq.~(\ref{second-case-eq-chi})]. At the end of inflation $l_{\theta} = e^{-N} l_0$, with $N\sim 55$, and therefore, in order to avoid having modifications to the power spectrum coming from 
 non-commutative effects we must require
\be
l_{0}^{-1} < 10^{-2} e^{-3N/2} |n_s-1|^{-1/4} M_{pl},
\ee
which, after replacing $N = 55$ and the value of the spectral index, gives:
\be
l_{0}^{-1} < 10^{-20} {\rm GeV}.
\ee
Since eq.~(\ref{cond-desitter}) must be valid at any time during inflation, and in particular at the epoch $\eta_0$ where our reference perturbations are crossing the horizon, the previous bound additionally implies $H \ll 10^{-20} {\rm GeV}$. We therefore conclude that this type of non-commutativity is incompatible with slow-roll inflation as a mechanism to generate the anisotropies presently observed at the CMB.

In addition, in the present case as well, we see from (\ref{mfeq2}) that the curvature perturbations continue to be damped after horizon crossing until inflation ends. Hence we compute in an entirely analogous manner to (\ref{psc2}), the power spectrum observed in the CMB to be:

\eq{fcps}{P(k) = \frac{H^2}{2M^2_{pl}}\frac{1}{\epsilon}\frac{e^{-6N(k)}}{k^2}\theta H^4,}
where again we note an extremely damped, blue tilted spectrum ($e^{-6N(k)}/k^2 \to 0$ as $k\to 0$, since $N(k)$ is the number of e-folds inflation lasts after the mode $k$ crosses the horizon).

Hence long wavelength power is greatly suppressed by the stiffening of modes even in the case where the non-commutativity tensor is redshifted by the expansion of spacetime. This is to be viewed as an analogous situation to the discontinuity in $\theta$ found in other contexts (see discussion surrounding (\ref{bfncq})). 
In summary, we have seen how UV/IR mode mixing due to the underlying non-commutative structure of spacetime drastically modifies the predictions of inflation regardless of the metric dependences of the non-commutativity tensor. The stiffening of long wavelength modes is a well known feature of non-commutative theories \cite{gubser}, and here we have seen how it can nullify one of the more successful features of inflation.

\section{Conclusions}

Slow-roll inflation provides a powerful framework to address speculative physics at very high energies. In this paper we have investigated the particular case of non-commutative geometry and its implications for the generation of scalar perturbations during inflation. We have shown that the phenomenon of UV/IR mode mixing, arising from one-loop corrections to the inflaton quadratic action, affects the evolution of fluctuations \mbox{--even} after horizon crossing-- in an unacceptable way.

To make our study as general as possible, we have considered three different situations, characterized by different time evolutions of the non-commutative lengthscale $l_{\theta}$. Interestingly, in the cases where $\theta^{i}_{\,\, j}$ and $\theta_{ij}$ are constant, we find that counter to the intuition that during inflation, the rapidly decaying theta tensor in the right hand side of eq.~(\ref{spat-non-commutativity}) would render any effects of non-commutativity null, we find the effects of UV/IR mode mixing to be {\it even more pronounced}. The reason for this is well understood, and comes down to the fact that many features of non-commutative theories are different in the $\theta\to 0$ limit than if there was no commutativity to begin with, such as various amplitudes \cite{min} and the phase structure of the theory \cite{gubser} among others. A rather stark demonstration of this comes from when one considers the beta function for non-commutative QED \cite{martin}, 
\eq{bfncq}{\beta = -\frac{g^3}{16\pi^2}\Bigl(\frac{22}{3} - \frac{4}{3}N_{\rm f} \Bigr),}
with $N_{\rm f}$ the number of Dirac fermion species. The new negative term arises from non-commutative photon self interactions, and is independent of $\theta$, provided non-commutativity exists in the first place. In an analogous manner, we find that even if the theta tensor is being rapidly redshifted, the consequence for the CMB two point function is rather pronounced. 

We thus conclude, echoing the sentiment expressed in the preliminaries, that as far as very high energy physics is concerned, the phenomenon of UV/IR mode mixing offers us a complimentary lever (on high energy physics) to the effects of inflation blowing up modes. However as we have seen in this context, the effects are so pronounced and in conflict with observations as to strongly disfavour the underlying model with manifests this mode mixing.

\section{Acknowledgements}
S.P. wishes to thank the Lorentz Institute at Leiden University, and the CCPP at NYU, and Gregory Gabadadze in particular for hospitality during the preparation of this report. G.A.P. wishes to thank the theory group at the Humboldt University for similar hospitality. G.A.P. is supported by The Netherlands Organization for Scientific Research (N.W.O.) under the VICI and VIDI programmes. S.P. is supported at the Humboldt University with funds from the Volkswagon foundation, and is grateful to Jan Plefka for this.

\begin{appendix}

\section{The star product}  \label{app-A}

The purpose of this appendix is to demonstrate why (\ref{sp}) represents the structure relations (\ref{ncg}) on function space. Starting with relations
\eq{rcomm}{[\hat x^\mu,\hat x^\nu] = i\theta^{\mu\nu},}
where each $\hat x^\mu$ is a Hermitean operator, we define the bounded operators
\eq{opdef}{T(k) = e^{ik_\mu \hat x^\mu}.}
These operators satisfy the relations:
\bea
T^{+}(k) &=& T(-k),  \label{rel1} \\
T(k)T(k') &=& T(k + k')e^{\frac{-i}{2}k_\mu k'_\nu\theta^{\mu\nu}},  \label{rel2}\\
tr T(k) &=& \Pi_{\mu}\delta(k_\mu),  \label{rel3}
\eea
where (\ref{rel2}) follows from the Campbell-Baker-Hausdorf relations, and the trace in (\ref{rel3}) is taken over the appropriate Hilbert space spanned by the representation (in 2-d, it would correspond to the usual Hilbert space used to represent the canonical pair $q = \hat x^1$ and $p = \hat x^2$, where $[q,p] = i$). Next, we associate to any classical function $\phi(x)$, the operator $\Phi$

\begin{eqnarray*}
\label{ope}
\Phi &=& \frac{1}{2\pi^n}\int d^nx d^nk ~ T(k) e^{-ik_\mu x^\mu}\phi(x)\\
&=& \frac{1}{2\pi^{n/2}}\int d^nk ~T(k) \tilde\phi(k), 
\end{eqnarray*}

\noindent where $\tilde\phi(k)$ is the Fourier transform of the function $\phi(x)$. In this sense, we can consider the operator $\Phi$ as the object that would result from Fourier transforming back from $\tilde\phi(k)$, except that we convolute with (\ref{opdef}) instead of the usual c-number phase. Using the trace relation (\ref{rel3}), we can recover $\phi(x)$ from $\Phi$:

\eq{pinv}{\phi(x) = \frac{1}{(2\pi^{n/2})}\int d^nk e^{ik_\mu x^\mu} ~tr~\Phi T^+(k),}

\noindent so that we have established a one to one and onto map between the space of c-number functions and operators over some Hilbert space. This map effects the standard prescription of `putting hats' on the c-number functions:

\eq{stanp}{x^\mu \leftrightarrow \hat x^\mu.}
When products of $x^\mu$ occur that are ambiguous, the above map returns the `Weyl ordered' or totally symmetrized product of operators, hence this mapping is referred to as the Weyl correspondence. 

Now we know that on the Hilbert space side of this correspondence, there exists the usual algebra of linear operators. That is, any two operators $\Phi_1$ and $\Phi_2$ multiply as $\Phi_1\Phi_2$. We can then ask what c-number function this operator maps back onto, and denoting this function as $\phi_1*\phi_2(x)$, we find after some straightforward manipulations that: 

\begin{eqnarray*}
(\phi_1*\phi_2)(x) &=& \frac{1}{(2\pi^{n/2})}\int d^nk e^{ik_\mu x^\mu} ~tr~[\Phi_1\Phi_2 T^+(k)]\\
&=&  e^{\frac{i\theta^{\mu\nu}}{2}\frac{\partial}{\partial x^\mu}\frac{\partial}{\partial y^\nu}}\phi_1(x) \phi_2(y)|_{y=x} .
\end{eqnarray*}     

\noindent Hence in order for this one to one map between c-numbers and operators to be a homomorphism (such that we reproduce the operator algebra on the function space), we see that we have to deform the usual product into the so called star product as outlined above. In this way, we can encode the non-commutative spacetime structure implied by (\ref{rcomm}) on our function space by simply replacing all commutative products with the star product (\ref{sp}). 

\section{Loop corrections} \label{app-B}

In this appendix, we derive the relevant quadratic terms of the effective action for cosmological perturbations in the presence of spatial non-commutativity. As we have seen in the preliminaries for this report, loop effects can drastically modify the structure of the quadratic part of the one loop effective action at low momenta relative to the commutative case. We shall demonstrate a similar phenomenon for cosmological perturbations {\it even if loop effects are insignificant in the corresponding commutative situation}. In order to calculate this effect, we refer to the formalism and results of \cite{jm, sloth}. We begin by considering the action for the metric perturbations up to third and fourth order, after which we consider which terms contribute the leading contributions to the (quadratic) one loop effective action. Since we are interested in computing corrections to the CMB observables that arise from non-planar UV/IR mixing, we focus on the terms which are leadingly singular in the IR, as it is these term which determines the leading corrections to the mode functions that would imprint themselves on the CMB.

Following \cite{jm, sloth}, we begin with the ADM decomposition 
\eq{adm}{ds^2 = -N^2dt^2 + h_{ij}(dx^i + N^idt)(dx^j + N^jdt),}
and work in the uniform curvature gauge:
\eq{gfjm}{h_{ij}= a^{2}\delta_{ij},~\phi = \phi_0 + \varphi,~ \mathcal N = 1+\alpha,~\mathcal N^i = \partial_i\chi,}
We start with the action:
\eq{saction}{S = \frac{M^2_{pl}}{2}\int \sqrt{-g}[R - (\partial\phi)^2 - 2M^2_{pl}V(\phi)],}
Where we note that in this convention, the field $\phi$ is dimensionless. Upon expanding the Einstein-Hilbert action coupled to the inflaton field to first order, the shift and the lapse functions $N$ and $N^i$ turn out to be Lagrange multipliers. Writing $\alpha = \alpha_1 + \alpha_2 +...$ and $\chi = \chi_1 + \chi_2 + ...$ where the subscripts denote the order of the perturbation, we find that:
\begin{eqnarray}
\alpha_1 &=& \frac{1}{2}\frac{\dot\phi_0}{H}\varphi \\
\label{chi}
\partial^2 \chi_1 &=& -\frac{1}{2}\frac{\dot\phi_0}{H}\dot{\varphi} -\frac{1}{2}\frac{\dot\phi_0}{H}\frac{\dot H}{H}\varphi + \frac{1}{2}\frac{\ddot\phi_0}{H}\varphi
\end{eqnarray}
which when solved for and plugged back into the action, results (after many integrations by parts) in \cite{jm, sloth}
\eq{pertact}{S = -\frac{M^2_{pl}}{2}\int \sqrt{-g^0}~[ (\partial \varphi)^2 +m_0^2\varphi^2],}
where $m^2 \equiv V''M^2_{pl} - 6\epsilon H^2$ and the background metric $g^0_{\mu\nu}$ is given by the associated line element
\eq{bgm}{ds^2 = -dt^2 + a^{2}dx^idx^i,}
and the slow roll parameter $\epsilon$ has been defined as
\eq{defsr}{\epsilon = \frac{\dot\phi_0^2}{2 H^2}.}
[Recall that now $M_{pl}$ appears explicitly in the definition of the scalar potential in eq.~(\ref{saction})]. The action to third order is found to be: 
\begin{eqnarray}
\nonumber
S_3 &=& M^2_{pl}\int \!\! \sqrt{-g^0}\Bigl[ -\frac{1}{4}\frac{\dot\phi_0}{H}\dot\varphi^2\varphi - \frac{1}{4a^2}\frac{\dot\phi_0}{H}\varphi(\partial\varphi)^2 \nonumber\\
&& 
- \dot\varphi\partial_i\chi_1\partial_i\varphi  
+ \Bigl(\frac{3}{8}\frac{\dot\phi_0^3}{H} - \frac{1}{8}\frac{\dot\phi_0^3}{H^2}\dot H + \frac{1}{8}\frac{\dot\phi_0^2}{H}\ddot \phi_0 \nonumber\\ 
&& 
- \frac{M^2_{pl}}{4}\frac{\dot\phi_0}{H}V'' - \frac{M^2_{pl}}{6}V'''\Bigr)\varphi^3   
+ \frac{1}{8}\frac{\dot\phi_0^3}{H^2}\dot\varphi\varphi^2 \nonumber\\
&&
+\frac{1}{4}\frac{\dot\phi_0}{H}\varphi\Bigl(\partial^2\chi_1\partial^2\chi_1 - \partial_i\partial_j\chi_1\partial_i\partial_j\chi_1 \Bigr) \Bigr] .
\label{3pa}
\end{eqnarray}
This is expression (3.9) in \cite{jm} after we employ the field redefinition (3.10) in the same reference. We note from (\ref{chi}) that $\chi_1$ has mass dimension $-1$. In the non-commutative case, where the non-commutativity is a constant purely spatial tensor (so integrating by parts over time yields no extra terms), we simply replace all products in the above with the star product (\ref{sp}), as inflaton field fluctuations feel this non-commutativity being the modulus parametrizing the locus of open string endpoints. Background quantities do not feel this non-commutativity as it is purely spatial. In the resulting expression, we would like to extract the terms that contributes to the leading divergence in the IR for the quadratic part of the one loop effective action. As seen in the preliminaries, non-planar cubic vertices contribute at most logarithmic IR divergences, unless there are derivatives acting on the vertex. To this end, we begin with the first term in the above as it has the most derivatives that could run along a loop (recalling from (\ref{chi}) that $\chi \sim \partial^{-2}\partial_t\varphi$) to leading order in $\epsilon$ (the only other terms which have as many derivatives are subleading in $\epsilon$). That is, we consider the interaction vertex \eq{intvert}{S_3 = \frac{M^2_{pl}}
 {2\sqrt 2}\int \sqrt{-g^0}~\sqrt\epsilon~\varphi\nabla_\mu\varphi\nabla_\nu\varphi~\bar g^{\mu\nu},}
where $\bar g_{\mu\nu}$ is the metric tensor multiplied by the matrix $diag (-1,1,1,1)$. However, it is easily seen that integrating this term by parts yields terms proportional to $\sim m^2\varphi^3$, $\sim \dot\varphi\varphi^2$ and terms of fourth order. Hence at third order, we only have to consider terms formally corresponding to the to the last three lines of (\ref{3pa}). To leading order, these terms will only contribute logarithmic IR divergences, as can also be seen by way of the field redefinition:
\eq{fred}{\varphi = \varphi_c - \frac{\ddot\phi_0}{\dot\phi_0}\varphi_c^2 -\frac{1}{8}\frac{\dot\phi_0}{H}\varphi_c^2 - \frac{\dot\phi_0}{4H}\partial^{-2}[\varphi_c\partial^2\varphi_c] +...}
As shown in \cite{jm}, in terms of $\varphi_c$, (\ref{3pa}) becomes: 
\eq{3par}{S_3 = M^2_{pl}\int \sqrt{-g}\sqrt\epsilon H \dot\varphi^2_c\partial^{-2}\dot\varphi_c.}
Given that we need two vertices of this interaction to construct the non-planar diagram in fig 2, with the relevant diagrams involving only time derivatives acting on the external legs, we see that this diagram yield at most logarithmically IR divergent contributions. 
At fourth order, we consider the action for $\varphi$ as computed by \cite{sloth}, where through similar considerations as above, we extract the leading relevant terms:
\begin{eqnarray}
\label{s4r}S_4 = &-&M^2_{pl}\int\sqrt{-g}\Bigl[\frac{15}{64}\frac{\dot\phi_0^4}{H^2} + \frac{1}{16}\frac{\dot\phi_0^2}{H^2}V''\\ \nonumber &+& \frac{1}{12}\frac{\dot\phi_0}{H}V''' + \frac{1}{24}V'''' \Bigr]\varphi^4 + ...      
\end{eqnarray} 
We symbolically denote the above interaction vertex as
\eq{foa}{H^2M^2_{pl} \delta^2\varphi^4,}
where $\delta$ is taken to be commensurate to the slow roll parameters $\epsilon$, $\eta$, etc..., and its precise value will depend on the details of the background solution. We now realize that non-planar one loop diagrams such as those considered in fig 1 are generated by this vertex. This will contribute a non-local counter term of order $\delta^2 H^2/M^2_{pl}$ from the loop integral:
\eq{liquart}{\delta^2 H^2\int \frac{d^4k}{k^2 + m^2}e^{ik\times p},}
with the resulting one loop effective action for the perturbations (up to a factor of order unity $c$, which depends on the precise details of the background dynamics):   
\eq{pertacts}{S =  \frac{M^2_{pl}}{2}\int \sqrt{-g}~d^4x \Bigl[ \varphi \square\varphi + m^2\varphi^2 + \frac{\delta^2H^2}{M^2_{pl}}\varphi\frac{c}{\partial\circ\partial}\varphi\Bigr].}
Converting the above expression into the action for the curvature perturbation $\zeta$ using the relation \cite{rhb, jm} (which is valid at the quadratic level in cour case as well as the non-commutativty tensor does not contract any time derivatives):
\eq{mukh}{\varphi = -\sqrt{2\epsilon}\zeta,}
we finally obtain the quadratic action for the curvature perturbation as:
\eq{pertactss}{S = M^2_{pl}\int d^4x\sqrt{-g^0}\epsilon \Bigl[\zeta\square\zeta +  \zeta\frac{\lambda}{\partial\circ\partial}\zeta\Bigr],}
where
\be
\lambda = \frac{\delta^2 H^2 c}{M^{2}_{pl}} . 
\ee
To finish, we notice that disregarding the last two terms in eq.~(\ref{s4r}), we are left with $\lambda \sim H^2 \epsilon (\eta + 7 \epsilon/2)/M_{pl}^{2}$.

\end{appendix}

\end{document}